\documentclass[prb,floatfix,twocolumn,showpacs,amsmath,amssymb]{revtex4}
\usepackage{graphicx}
\usepackage{epstopdf}
\usepackage{epsfig}
\usepackage{dcolumn}
\usepackage{bm}
\usepackage[colorlinks=true,linkcolor=blue,citecolor=blue]{hyperref}

\begin{document}
\title{Variational Monte Carlo study of gapless spin liquid in the spin-$1/2$ XXZ antiferromagnetic model on the kagome lattice}
\author{Wen-Jun Hu$^1$, Shou-Shu Gong$^1$, Federico Becca$^2$, and D. N. Sheng$^1$}
\affiliation{
$^1$ Department of Physics and Astronomy, California State University, Northridge, California 91330, USA \\
$^2$ Democritos National Simulation Center, Istituto Officina dei Materiali del CNR, 
and SISSA-International School for Advanced Studies, Via Bonomea 265, I-34136 Trieste, Italy
}

\begin{abstract}
By using the variational Monte Carlo technique, we study the spin-$1/2$ XXZ antiferromagnetic 
model (with easy-plane anisotropy) on the kagome lattice. A class of Gutzwiller projected fermionic 
states with a spin Jastrow factor is considered to describe either spin liquids (with $U(1)$ 
or $Z_2$ symmetry) or magnetically ordered phases (with ${\bf q}=(0,0)$ or ${\bf q}=(4\pi/3,0)$). 
We find that the magnetic states are not stable in the thermodynamic limit. Moreover, there is no 
energy gain to break the gauge symmetry from $U(1)$ to $Z_2$ within the spin-liquid states, as 
previously found in the Heisenberg model. The best variational wave function is therefore the 
$U(1)$ Dirac state, supplemented by the spin Jastrow factor. Furthermore, a vanishing $S=2$ spin gap 
is obtained at the variational level, in the whole regime from the $XY$ to the Heisenberg model.
\end{abstract}

\pacs{75.10.Jm, 75.10.Kt, 75.40.Mg, 75.50.Ee}

\maketitle

{\it Introduction}. 
Quantum spin liquids with topological order and fractional excitations are exotic states 
of matter that do not show any local order down to zero temperature.~\cite{Balents2010} 
Their importance for the field of correlated systems is directly related to the connection 
to unconventional electron pairing, thus giving a clue to explain the mechanism of 
high-temperature superconductivity.~\cite{Anderson1973,Lee2006} In the last two decades, 
there have been intensive studies suggesting that quantum spin liquids might be stabilized 
at low temperatures in realistic two-dimensional frustrated magnetic systems. The spin-$1/2$ 
Heisenberg antiferromagnetic model on the kagome lattice represents one of the most promising 
examples.~\cite{Lee2008} From the experimental side, the so-called Herbertsmithite shows 
very promising signatures for magnetically disordered phases down to extremely low 
temperatures.~\cite{mendels2007,helton2007,vries2009,han2012} Even more interestingly, many 
experimental probes suggested the existence of gapless spin excitations; in particular, 
neutron scattering measurements highlighted the presence of a broad continuum of excitations 
down to small energies.~\cite{vries2009,han2012}

For Herbertsmithite, it is widely believed that the gross features can be captured by the 
spin-$1/2$ Heisenberg antiferromagnet on the kagome lattice with the nearest-neighbor interactions 
only. This model has been studied by several analytical and numerical approaches in recent years,
with contradicting outcomes.~\cite{Wang2006,Ran2007,Iqbal2011,Singh2007,Singh2008,Jiang2008,Evenbly2010,Yan2011,Depenbrock2012,nakano2011,Gotze2011,Jiang2012nature,Messio2012,Xie2014,Lu2014unification}
In particular, accurate density-matrix renormalization group (DMRG) calculations highlighted the
possibility that the ground state can be a fully gapped $Z_2$ topological spin liquid.~\cite{Yan2011,Depenbrock2012}
A different scenario has been put forward by using variational Monte Carlo approaches based upon 
Gutzwiller projected fermionic state,~\cite{Ran2007,Iqbal2011,Hermele2008} which find a gapless 
spin liquid with a competing ground-state energy.~\cite{Iqbal2013,Iqbal2014}

In order to clarify the nature of the spin-liquid phase, several authors considered the effect 
of different ``perturbations'' to the nearest-neighbor Heisenberg model, the most obvious one 
being a second-neighbor super-exchange.~\cite{Tay2011,aps_white,Jiang2012nature,Iqbal2015,ssgong2015,Kolley2015} 
However, given the lack of the consistent results,~\cite{aps_white,Jiang2012nature,ssgong2015,Kolley2015}
it is still not clear if this term helps the stabilization of the spin liquid or not.
The inclusion of an additional third-neighbor couplings~\cite{Messio2012,gong2014kagome,he2014csl,ssgong2015} 
or three-spin chiral interactions~\cite{bauer2014} stabilizes a topological spin liquid with spontaneously
time-reversal symmetry breaking,~\cite{kalmeyer1987,wen1989csl} which has been identified as the 
$\nu=1/2$ bosonic quantum Hall state.~\cite{gong2014kagome,he2014csl,ssgong2015} Recent DMRG 
studies found that this chiral state can persist also by changing the magnetic anisotropy within 
the XXZ model.~\cite{He2015,Wei2014} In this respect, the XXZ model with the only nearest-neighbor 
interactions has not been thoroughly investigated. Some recent calculations have pointed out the 
possibility that different magnetic orders are favored for the $XY$ and Heisenberg models, based upon 
order-by-disorder mechanisms.~\cite{Chernyshev2014,Gotze2015} This situation should take place for 
large enough spin $S$, while for $S=1/2$ magnetically disordered states should be expected. Indeed, 
DMRG calculations have suggested the existence of a spin-liquid phase; however, it remains unclear 
if there is a phase transition between the $XY$ and the Heisenberg models for $S=1/2$.~\cite{He2015,Wei2014}
In particular, the $XY$ model could have a vanishing-small spin gap, which is compatible with a 
gapless quantum spin liquid in the thermodynamic limit.~\cite{Wei2014} Therefore, the spin 
anisotropy in the nearest-neighbor coupling represents a very promising way to unveil the nature 
of the spin-liquid phase of the Heisenberg model.

In this paper, we consider the following Hamiltonian:
\begin{equation}\label{eq:ham1}
{\cal H} = J_{xy}\sum_{\langle ij\rangle}(S^{x}_{i}S^{x}_{j}+S^{y}_{i}S^{y}_{j})+
           J_{z}\sum_{\langle ij\rangle}S^{z}_{i}S^{z}_{j},
\end{equation}
where $\langle ij \rangle$ denotes the sum over the nearest-neighbor pairs of sites, and 
${\bf S}_{i}=(S^{x}_{i}, S^{y}_{i}, S^{z}_{i})$ is spin-$1/2$ operator at each site $i$. 
In the following, we will set $J_{xy}=1$ as the energy scale. When $J_{z}=1$ ($J_{z}=0$), 
Eq.~(\ref{eq:ham1}) reduces to the Heisenberg ($XY$) model. Here, we focus on the region with 
$0 \le J_{z} < 1$, and study the stability of different variational wave functions including the $U(1)$ 
and $Z_2$ spin liquids, as well as the magnetic ordered states (the isotropic Heisenberg model
with $J_{z}=1$ has been thoroughly investigated in previous works~\cite{Iqbal2011,Iqbal2013,Iqbal2014}). 
The main results can be summarized as follow: by applying an energy optimization, there is no signal for 
the stabilization of a gapped spin liquid; also the inclusion of magnetic orders does not improve the 
$U(1)$ Dirac state. Instead, some energy gain can be obtained by including a (short-range) Jastrow factor.
Then, we construct the $S=2$ state by exciting 4 spinons in the $U(1)$ Dirac spin liquid and calculate 
the $S=2$ spin gap at different values of $J_{z}$. The variational results are compatible with the 
conclusion that the same gapless spin liquid persists from $J_{z}=1$ to $J_{z}=0$.
\begin{figure}
\includegraphics[width=0.7\columnwidth]{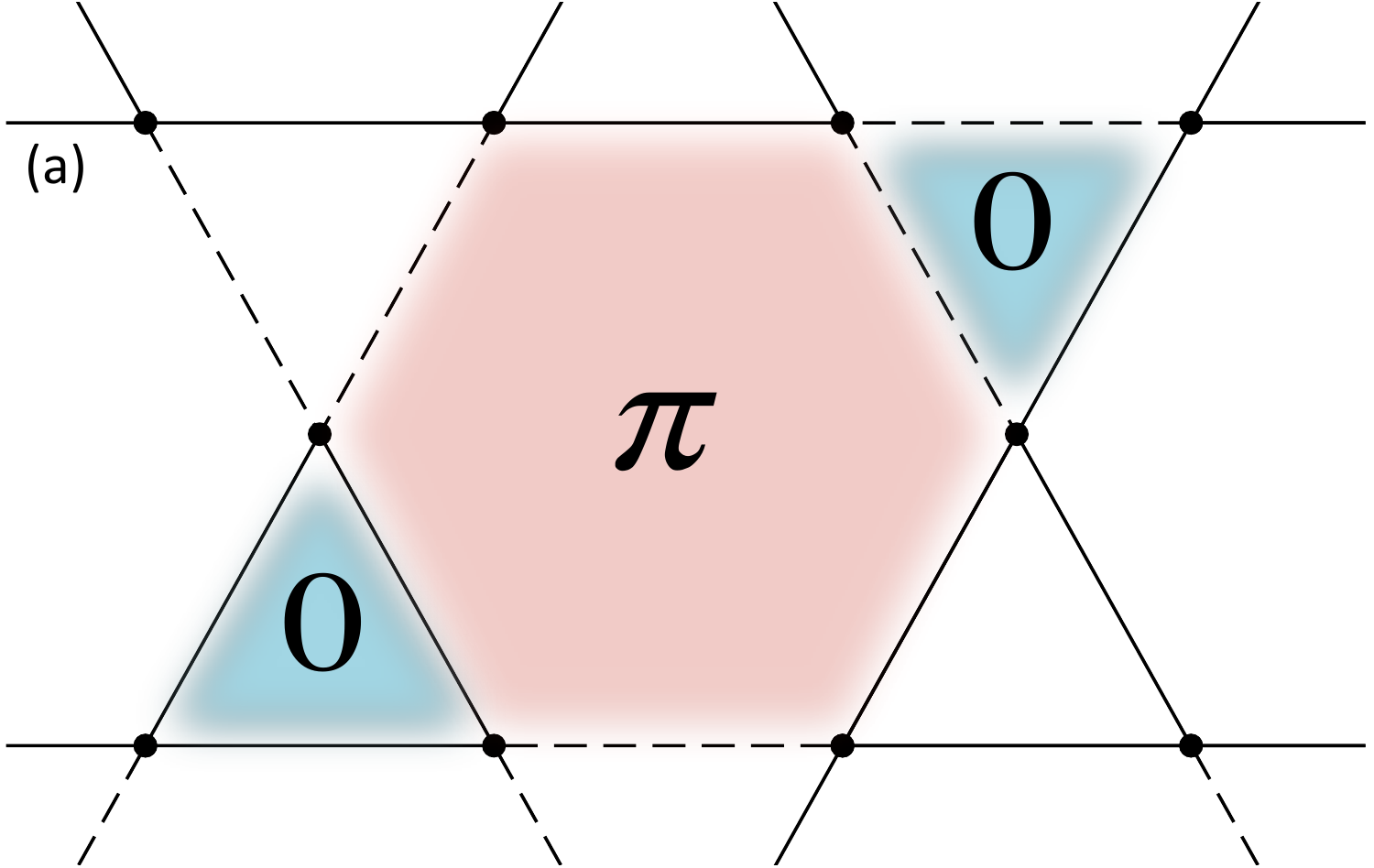}\quad
\includegraphics[width=0.7\columnwidth]{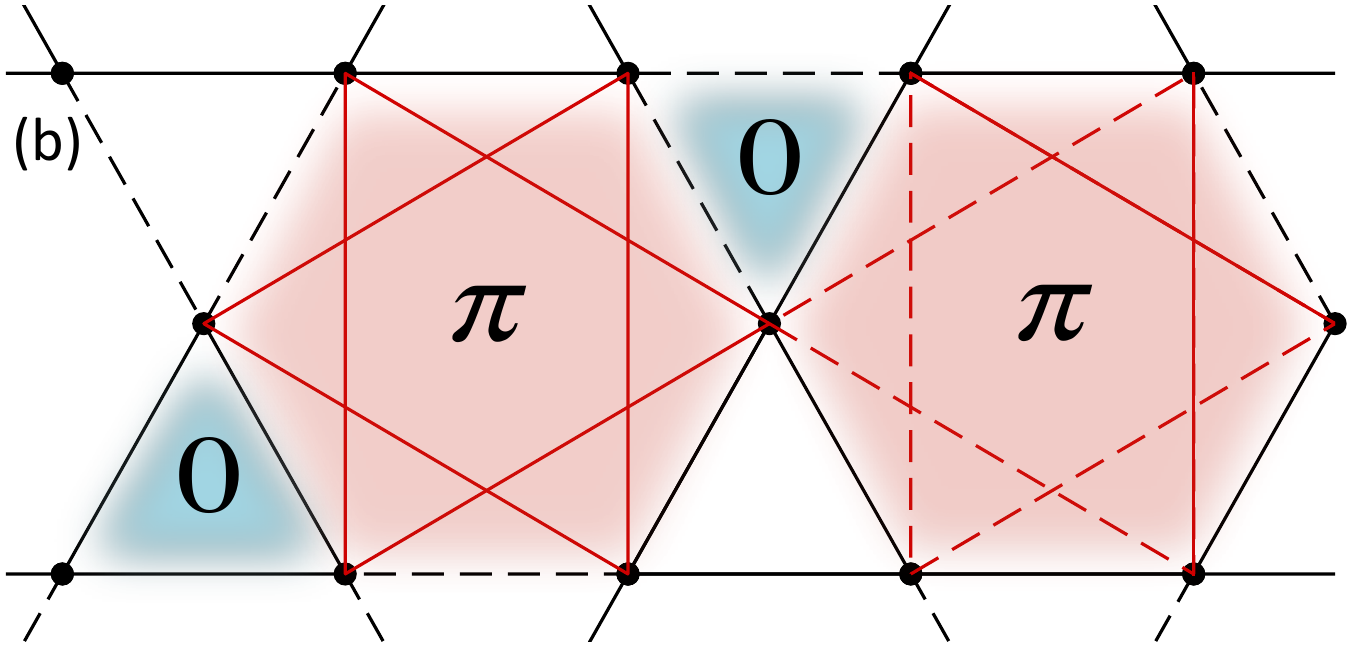}
\caption{\label{fig:wf}
(Color online) (a) The mean-field {\it Ansatz} of the $U(1)$ Dirac state: black solid (dashed) 
bonds denote positive (negative) hoppings ($t_1$). The unit cell is doubled to accommodate a 
magnetic flux $\Phi=\pi$ across hexagons and $\Phi=0$ across triangles. The same hopping amplitudes 
are also used to define the magnetic wave function obtained from Eq.~(\ref{eq:AF}).
(b) The mean-field {\it Ansatz} of the $Z_2[0,\pi]\beta$ state: red solid (dashed) lines 
indicate positive (negative) next-nearest-neighbor hoppings ($t_2$) and pairings ($\Delta_2$).}
\end{figure}

{\it Method and variational wave functions}.
Our variational wave functions are defined as
\begin{equation}\label{eq:vwf}
|\Psi_{v}\rangle=\mathcal{J}_{s}\mathcal{P}_{G}|\Psi_{0}\rangle,
\end{equation}
where $|\Psi_{0}\rangle$ is an uncorrelated wave function that is obtained as the ground state 
of an auxiliary Hamiltonian (see below); $\mathcal{P}_{G}=\prod_{i}(1-n_{i\uparrow}n_{i\downarrow})$ 
is the Gutzwiller projector, which enforces no double occupation on each site; 
$\mathcal{J}_{s}={\rm exp}(1/2 \sum_{ij}v_{ij}S^{z}_{i}S^{z}_{j})$ is the spin Jastrow factor,
$v_{ij}$ being variational parameters that depend upon the distance between sites $i$ and $j$.
We would like to stress the fact that such a Jastrow term, which includes the $z$ components of the 
spin operator, does not break any symmetry of the spin Hamiltonian in the easy-plane limit 
($J_{z}<1$), while it breaks the spin $SU(2)$ symmetry for the Heisenberg model ($J_{z}=1$). 
Here, we consider two cases for the auxiliary (non-interacting) Hamiltonian that are suitable 
for magnetic and spin-liquid wave functions. 

\begin{figure}[b]
\includegraphics[width=\columnwidth]{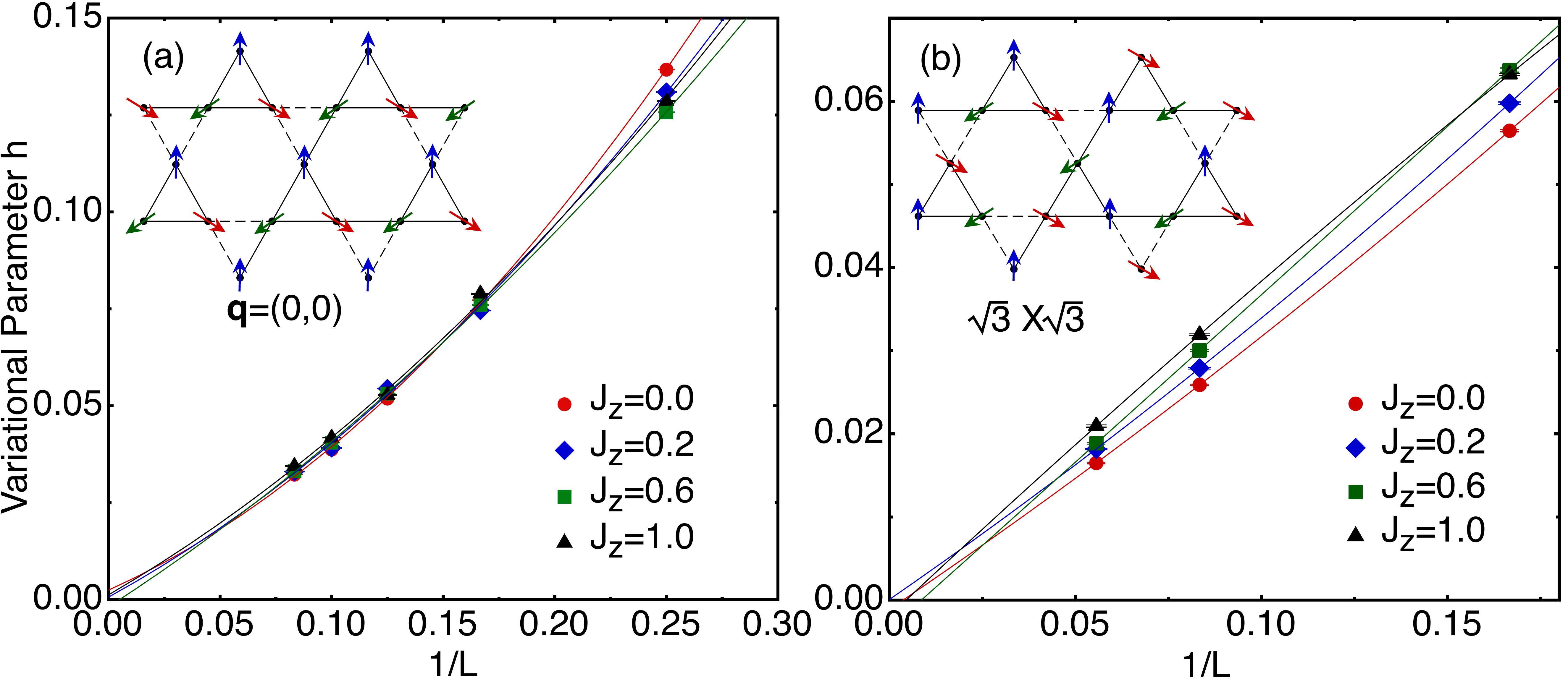}
\caption{\label{fig:afpara}
(Color online) The finite size scaling of the variational parameter $h$ as a function of the 
inverse geometrical diameter $1/L$ at different values of $J_{z}$ for the magnetic order with 
${\bf q}=(0,0)$ (a) and the $\sqrt{3}\times\sqrt{3}$ order (b). The quadratic fitting is used for 
all cases. The {\it Ansatz} for the magnetic order with ${\bf q}=(0,0)$ and $\sqrt{3}\times\sqrt{3}$ 
are also shown.}
\end{figure}

Magnetic states are defined from:
\begin{equation}\label{eq:AF}
{\cal H}_{\rm MAG} = \sum_{(i,j),\sigma}t_{ij}c^{\dag}_{i,\sigma}c_{j,\sigma} 
                   + h \sum_{i} {\bf M}_{i} \cdot {\bf S}_{i},
\end{equation}
where $c^{\dag}_{i,\sigma}$ ($c_{i,\sigma}$) creates (destroys) one electron at site $i$ 
with spin $\sigma$. We find that the best projected state within this class of wave functions 
has non-trivial hopping amplitudes, which define a magnetic flux $\Phi=\pi$ across hexagons 
and $\Phi=0$ across triangles, see Fig.~\ref{fig:wf} (they are exactly the ones that define 
the $U(1)$ Dirac spin liquid in Ref.~\onlinecite{Ran2007}). The magnetic order is defined by 
the (variational) parameter $h$ and the vector ${\bf M}_{i}$ that defines the periodicity; 
here, we consider coplanar states and restrict ${\bf M}_{i}$ in the $XY$ plane, i.e., 
${\bf M}_{i}=(\cos({\bf r}_{i}\cdot {\bf q}+\eta_{i}),\sin({\bf r}_{i}\cdot {\bf q}+\eta_{i}),0)$
(${\bf q}$ is the pitch vector and $\eta_{i}$ is the phase shift for sites within the same 
unit cell). In the following, we consider two antiferromagnetic patterns with ${\bf q}=(0,0)$ 
and ${\bf q}=(4\pi/3,0)$ (corresponding to the $\sqrt{3}\times\sqrt{3}$ order), see the insets 
of Fig.~\ref{fig:afpara}. With ${\bf M}_{i}$ in the $XY$ plane, the spin Jastrow factor correctly 
describes the relevant spin fluctuations around the classical spin state.~\cite{Manousakis1991} 
We would like to emphasize that the existence of magnetic long-range order is directly related 
to the presence of a finite parameter $h$ in Eq.~(\ref{eq:AF}).

Instead, spin-liquid wave functions are defined from:
\begin{eqnarray}
{\cal H}_{SL} &=& \sum_{(i,j),\sigma}t_{ij}c^{\dag}_{i,\sigma}c_{j,\sigma}+
                  \sum_{(i,j)}[\Delta_{ij}c^{\dag}_{i,\uparrow}c^{\dag}_{j,\downarrow}+h.c.] \nonumber\\
              &+& \mu\sum_{i\sigma}c^{\dag}_{i,\sigma}c_{i,\sigma}+ 
                  \Delta_{0}\sum_{i}[c^{\dag}_{i,\uparrow}c^{\dag}_{i,\downarrow}+h.c.],
\label{eq:BCS}
\end{eqnarray}
where, in addition to the hopping terms, there is also a singlet pairing ($\Delta_{ij}=\Delta_{ji}$); 
the on-site paring $\Delta_{0}$ and the chemical potential $\mu$ are also considered. It is possible 
to show that many different spin liquids can be constructed, depending on the symmetries of $t_{ij}$ 
and $\Delta_{ij}$, which may have $U(1)$ or $Z_2$ gauge structure and gapped or gapless spinon 
spectrum.~\cite{Luyuanming2011} In this paper, we consider two kinds of spin liquids, namely
the gapless $U(1)$ Dirac state and the gapped $Z_{2}[0,\pi]\beta$ state, as shown in Fig.~\ref{fig:wf}.
We emphasize that, for $J_{z}<1$, a spin Jastrow factor can be also included, since the spin $SU(2)$ 
symmetry is explicitly broken by the XXZ Hamiltonian~(\ref{eq:ham1}).

\begin{figure}
\includegraphics[width=\columnwidth]{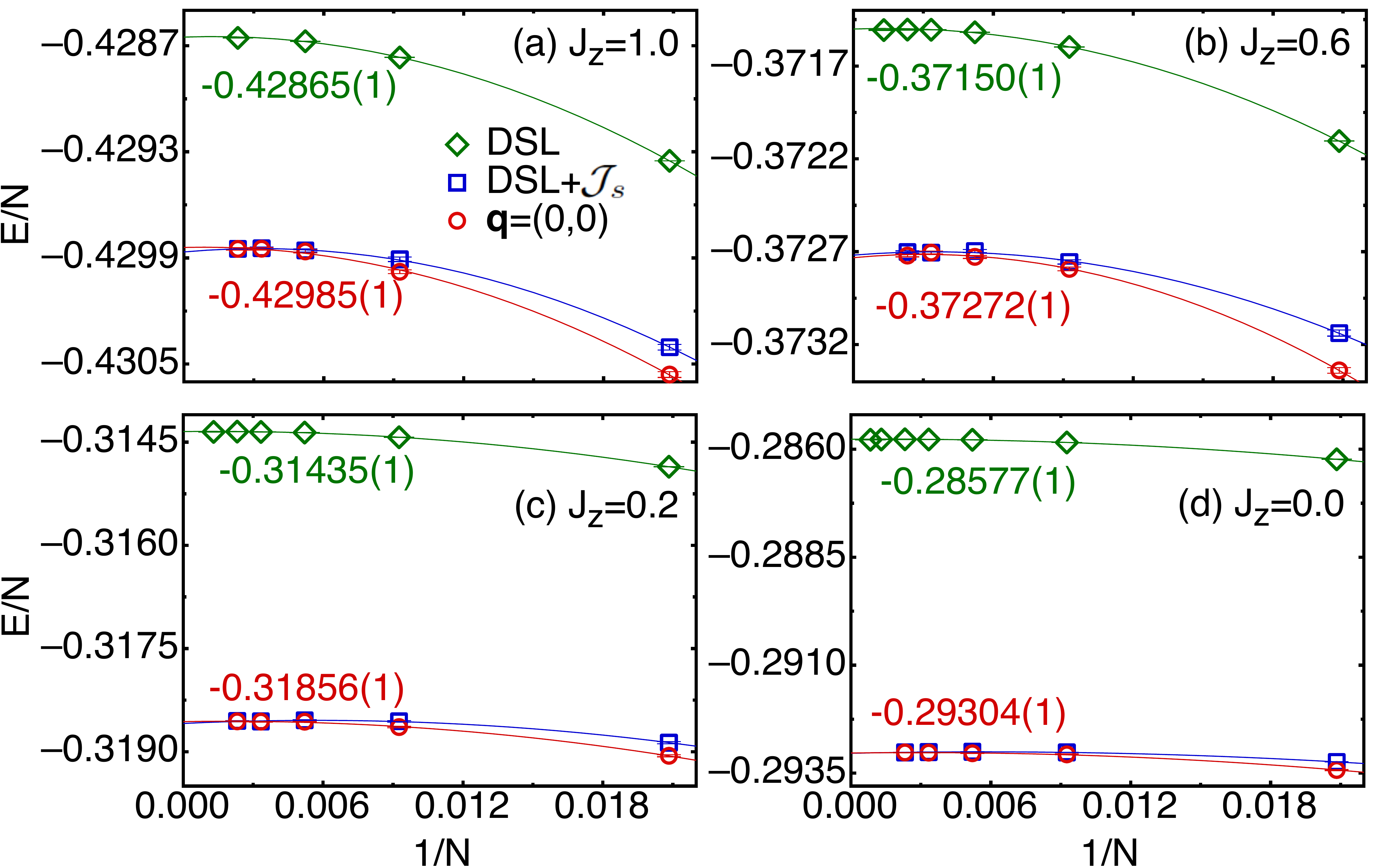}
\caption{\label{fig:energy}
(Color online) Energies per site of the $U(1)$ Dirac spin liquid (green diamonds) compared 
with the one of the magnetic state with ${\bf q}=(0,0)$ (red circles) for different values of
$J_{z}$. The $U(1)$ Dirac state with spin Jastrow factor is also reported (blue squares), 
which corresponds to the magnetic state with a vanishing variational parameter $h$.}
\end{figure}

In general, suitable boundary conditions in the auxiliary Hamiltonians are chosen, in order to have a 
unique mean-field ground state $|\Psi_{0}\rangle$. In order to fulfill the constraint of one electron
per site (imposed by the Gutzwiller projector) and to take into account the spin Jastrow factor, 
a Monte Carlo sampling is needed. Along the Markov chain that define the numerical simulation, only 
configurations belonging to the physical Hilbert space are proposed, so that the Gutzwiller projector is 
exactly implemented. To optimize the variational parameters in both the auxiliary Hamiltonians and the 
spin Jastrow factor, we use the stochastic reconfiguration (SR) optimization method to find the energetically 
favored state in variational Monte Carlo scheme.~\cite{Sorella2005} The SR optimization method allows us 
to perform the optimization with many variational parameters, and to obtain an extremely accurate 
determination of variational parameters. 

\begin{figure}[b]
\includegraphics[width=\columnwidth]{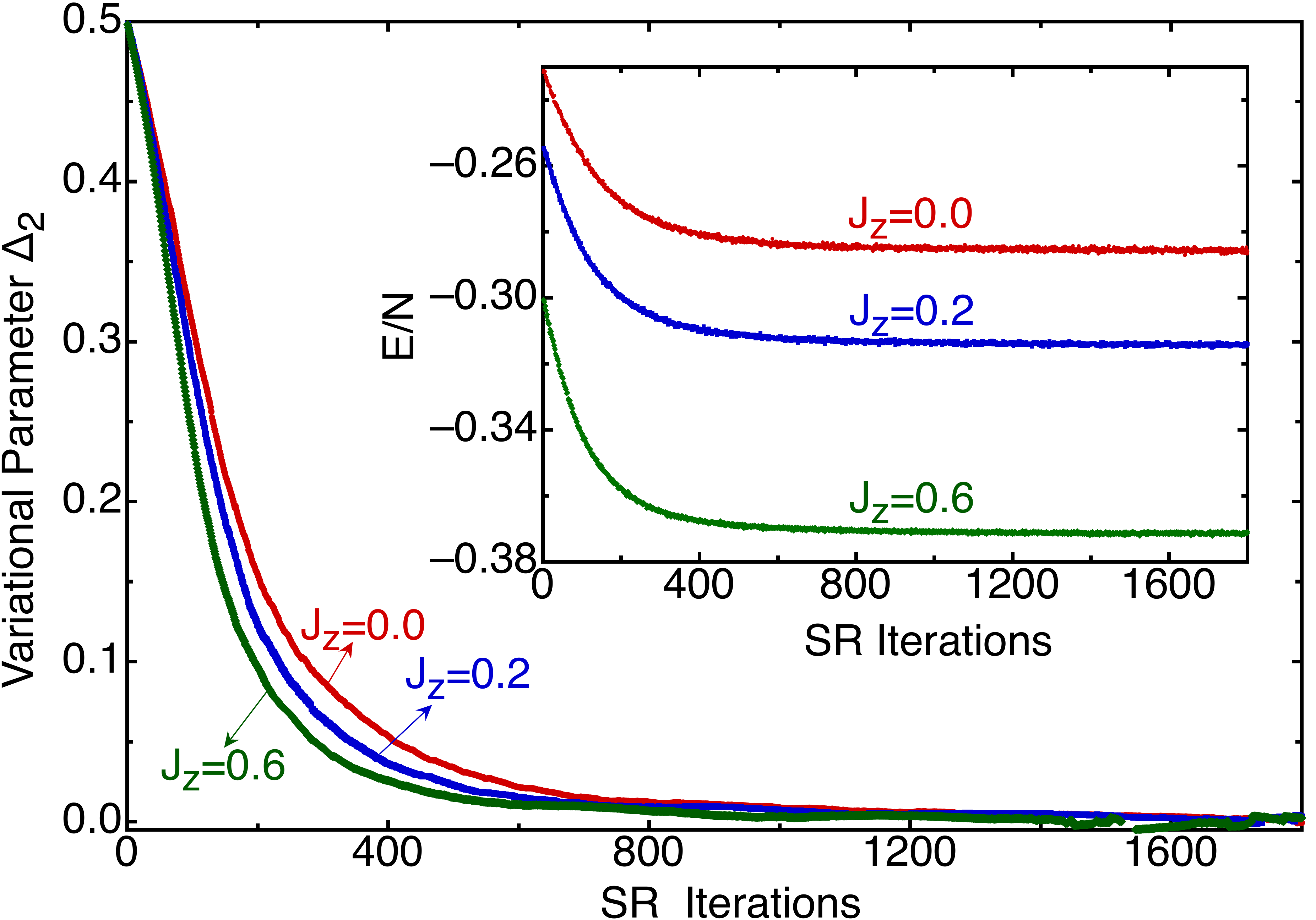}
\caption{\label{fig:z2beta}
(Color online) The variational parameter $\Delta_2$ for the $Z_2[0,\pi]\beta$ spin liquid 
for $J_{z}=0$, $0.2$, and $0.6$ on $L=16$ cluster. The variational energy per site as 
function of the SR iterations is shown in the inset.}
\end{figure}

{\it Results}.
We performed the variational Monte Carlo calculations on toric clusters with $L\times L\times3$ 
sites and periodic boundary conditions. Let us start with magnetic states. In Fig.~\ref{fig:afpara}, 
we show the size scaling of the magnetic order parameter $h$ of Eq.~(\ref{eq:AF}) for both
the states with ${\bf q}=(0,0)$ and $\sqrt{3}\times\sqrt{3}$ order. The latter one is not
frustrated by boundary conditions only when $L$ is a multiple of $3$. First of all, we find
that a finite magnetic parameter $h$ of Eq.(\ref{eq:AF}) can be stabilized on finite clusters, and
the two states have essentially the same energy within $10^{-4}J_{xy}$ for all the cases 
analyzed here. However, the most important outcome is that $h \to 0$ in the thermodynamic limit
(for both antiferromagnetic states), indicating that no magnetic order can be stabilized in
the XXZ model. Therefore, the auxiliary Hamiltonian from which the variational state is
constructed reduces to the $U(1)$ Dirac state of Ref.~\onlinecite{Ran2007}. Nevertheless, the 
parameters $v_{ij}$ of the spin Jastrow factor remain finite, with sizable values at short-range 
distances. This fact gives a non-negligible energy gain with respect to the $U(1)$ Dirac state, 
especially close to the $XY$ limit, see Fig.~\ref{fig:energy}. We stress that the presence of the 
spin Jastrow factor is just relevant to improve short-range observables, such as the energy.
We mention that the spin Jastrow factor gives a small energy gain of about $10^{-3}J_{xy}$ also in the 
Heisenberg case, see Fig.~\ref{fig:energy} (however, in this case, the spin Jastrow factor spoils the 
spin $SU(2)$ invariance of the Dirac spin liquid).

\begin{figure}
\includegraphics[width=\columnwidth]{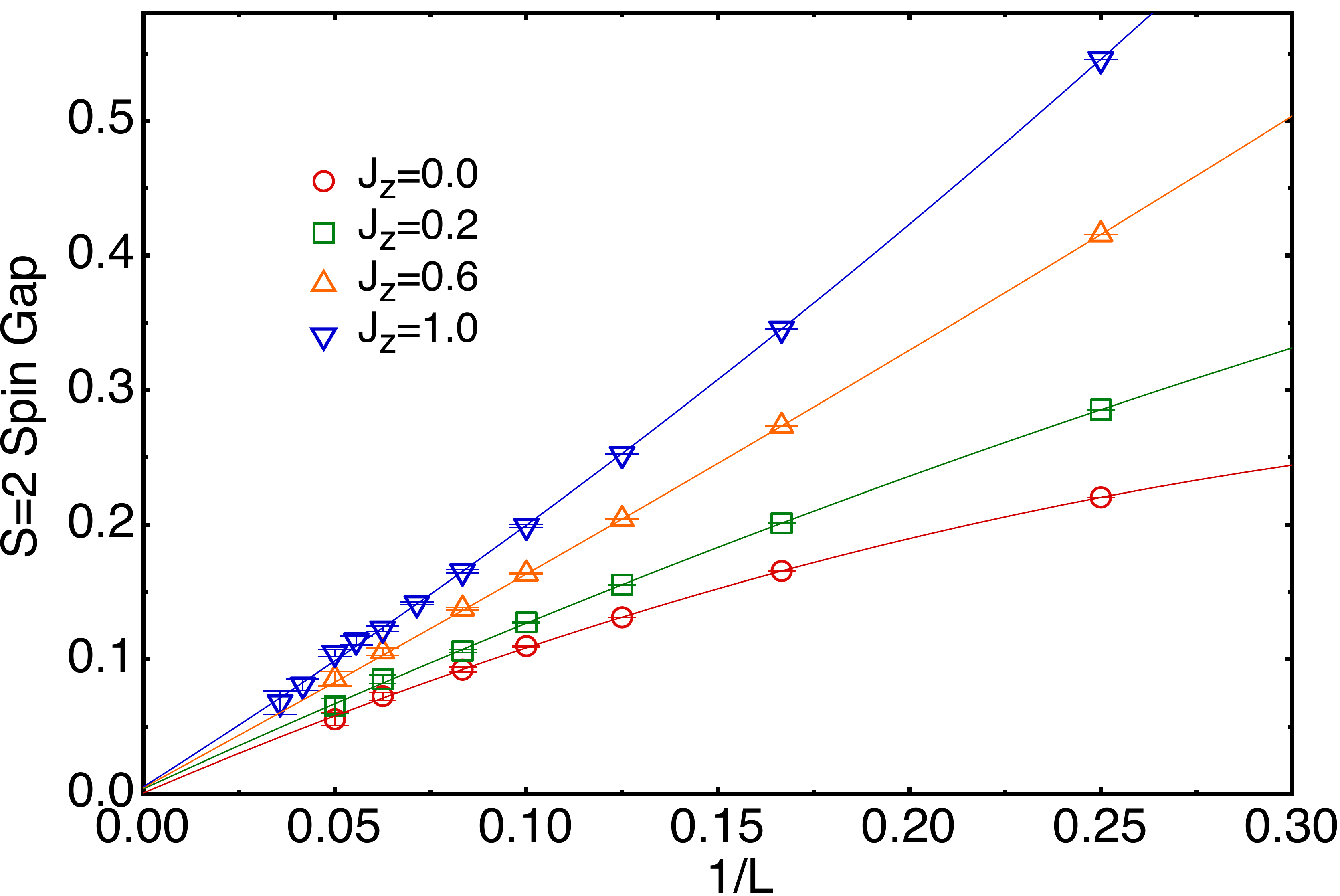}
\caption{\label{fig:spin2gap}
(Color online) The finite size scaling of the $S=2$ spin gap as a function of $1/L$ for
different values of $J_{z}$. The results for $J_{z}=1$ are from Ref.~\onlinecite{Iqbal2014}.}
\end{figure}

We now move to study the possible stabilization of $Z_2$ spin liquid states. According to the
classification of Ref.~\onlinecite{Luyuanming2011}, there is only one $Z_2$ gapped spin liquid that 
is directly connected to the $U(1)$ Dirac spin liquid, the so-called $Z_2[0,\pi]\beta$ state. 
In Ref.~\onlinecite{Iqbal2011}, the authors have shown that this gapped spin liquid cannot be stabilized 
in the Heisenberg model with $J_{z}=1$. Here, we would like to extend the analysis to the case of the 
XXZ model. The variational states is constructed from Eq.~(\ref{eq:BCS}), and the non-interacting wave 
function that defines the $Z_2[0,\pi]\beta$ state includes the nearest-neighbor $t_1$, the 
next-nearest-neighbor $t_2$ hoppings, the next-nearest-neighbor paring $\Delta_2$, which is responsible 
for the breaking from $U(1)$ to $Z_2$ symmetry, a chemical potential $\mu$, and the on-site pairing 
$\Delta_0$, see Fig.~\ref{fig:wf}. In the following, we do not consider the spin Jastrow factor, which 
may improve the energy but does not change the optimization of the variational parameter $\Delta_2$.
The optimization is shown in Fig.~\ref{fig:z2beta} for $J_{z}=0$, $0.2$, and $0.6$ on the $L=16$ cluster. 
The result is that, as for the Heisenberg case, both $\Delta_2$ and $\Delta_0$ (not shown here) go to 
zero for all the values of $J_{z}$ considered, even on finite clusters. The vanishing $Z_2$ parameters 
indicates that, similar to what has been found in the Heisenberg model,~\cite{Iqbal2011} the gapped 
$Z_2[0,\pi]\beta$ spin liquid is not stable in the XXZ model.

In summary, we obtain that the best variational state of the form~(\ref{eq:vwf}) that can be 
constructed from Eqs.~(\ref{eq:AF}) and~(\ref{eq:BCS}) is the $U(1)$ Dirac spin liquid, 
supplemented by a short-range spin Jastrow factor.~\cite{suppl} 

In the following, we compute the $S=2$ spin gap for the $U(1)$ Dirac state. The $S=2$ state is 
constructed by changing boundary conditions, in order to have 4 spinons in an 8-fold degenerate 
single-particle level at the chemical potential; a unique mean-field state is then obtained by taking 
all these spinons with the same spin. This $S=2$ state can be written in terms of a single determinant,
which is particularly easy to be treated within our Monte Carlo sampling. In the following, we do not 
consider the spin Jastrow factor, since its inclusion does not modify the qualitative picture.~\cite{suppl} 
Similarly to what has been done on the Heisenberg model,~\cite{Iqbal2014} we obtain the $S=2$ spin gap by 
computing separately the energies of the $S=0$ and $S=2$ states. In Fig.~\ref{fig:spin2gap}, we report 
the results for $J_{z}=0$, $0.2$, and $0.6$ (the case with $J_{z}=1$ from Ref.~\onlinecite{Iqbal2014} 
are also reported for comparison). First of all, we remark that, for each cluster size, the spin gap 
decreases by decreasing the value of $J_{z}$, indicating that the anisotropy in the spin super-exchange 
tends to close the finite-size gap. This result is in agreement with DMRG calculations in 
Ref.~\onlinecite{Wei2014}. Most importantly, the finite-size scaling with $L$ up to $20$ clearly indicates 
a vanishing spin gap for all values of $J_{z}$ considered here. Therefore, our analysis based upon 
Gutzwiller-projected states suggests that the same $U(1)$ Dirac state with gapless spinon excitations 
can be stabilized from $J_{z}=1$ to $J_{z}=0$.

{\it Conclusions}.
In summary, we investigated the XXZ model on the kagome lattice by using the variational Monte 
Carlo technique with the Gutzwiller projected fermionic states. We have studied different 
variational wave functions describing either magnetic states or spin-liquid phases. As previously 
obtained in the Heisenberg model,~\cite{Iqbal2011} the gapped $Z_2[0,\pi]\beta$ spin liquid 
cannot be stabilized for $J_{z}<1$, indicating a remarkable stability of the gapless $U(1)$ 
Dirac state. Moreover, the consideration of magnetic order does not give any energy gain in the 
thermodynamic limit, but a considerable energy gain can be obtained by the spin Jastrow factor. 
The $S=2$ spin gap, on any finite-size clusters, decreases with decreasing the value of $J_{z}$,
indicating that the best place to find a gapless spin liquid is most probably close to the
$XY$ limit. This outcome agrees with a recent DMRG study~\cite{Wei2014}, which suggested that
a critical state can be stabilized near the XY kagome model. In addition, 
we do not find any evidence for possible dimer states, as also suggested by DMRG 
calculations.~\cite{He2015,Wei2014}

Finally, we would like to mention that the application of few Lanczos steps to the $U(1)$ Dirac 
spin liquid, as already done in recent works for the Heisenberg model,~\cite{Iqbal2013,Iqbal2014} 
does not alter the results on the $S=2$ gap, although the large statistical errors for $L=8$
do not allow us to obtain as neat conclusions as in the Heisenberg model.~\cite{suppl}
In fact, even though the $U(1)$ Dirac spin liquid remains the best variational state within the
class of fermionic states that have been analyzed here, its accuracy slightly deteriorates
when decreasing the value of $J_{z}$, which makes the zero-variance extrapolation harder than
for the Heisenberg case.

{\it Acknowledgements}.
We thank Y. Iqbal for providing us with the variational data for $J_{z}=1$,
and thank W. Zhu for providing the DMRG data.
This research is supported by the National Science Foundation through grants DMR-1408560 
(W.-J.H, D.N.S) and PREM DMR-1205734 (S.S.G.), and by the Italian MIUR through PRIN 2010-11 (F.B.).

\bibliographystyle{apsrev}
\bibliography{kagome_xxz}{}

\clearpage
\begin{center}
\begin{large}
{\bf Supplemental Material}
\end{large}
\end{center}
{\it The spin Jastrow factor--}
\begin{figure}
\begin{center}
\includegraphics[width=\columnwidth]{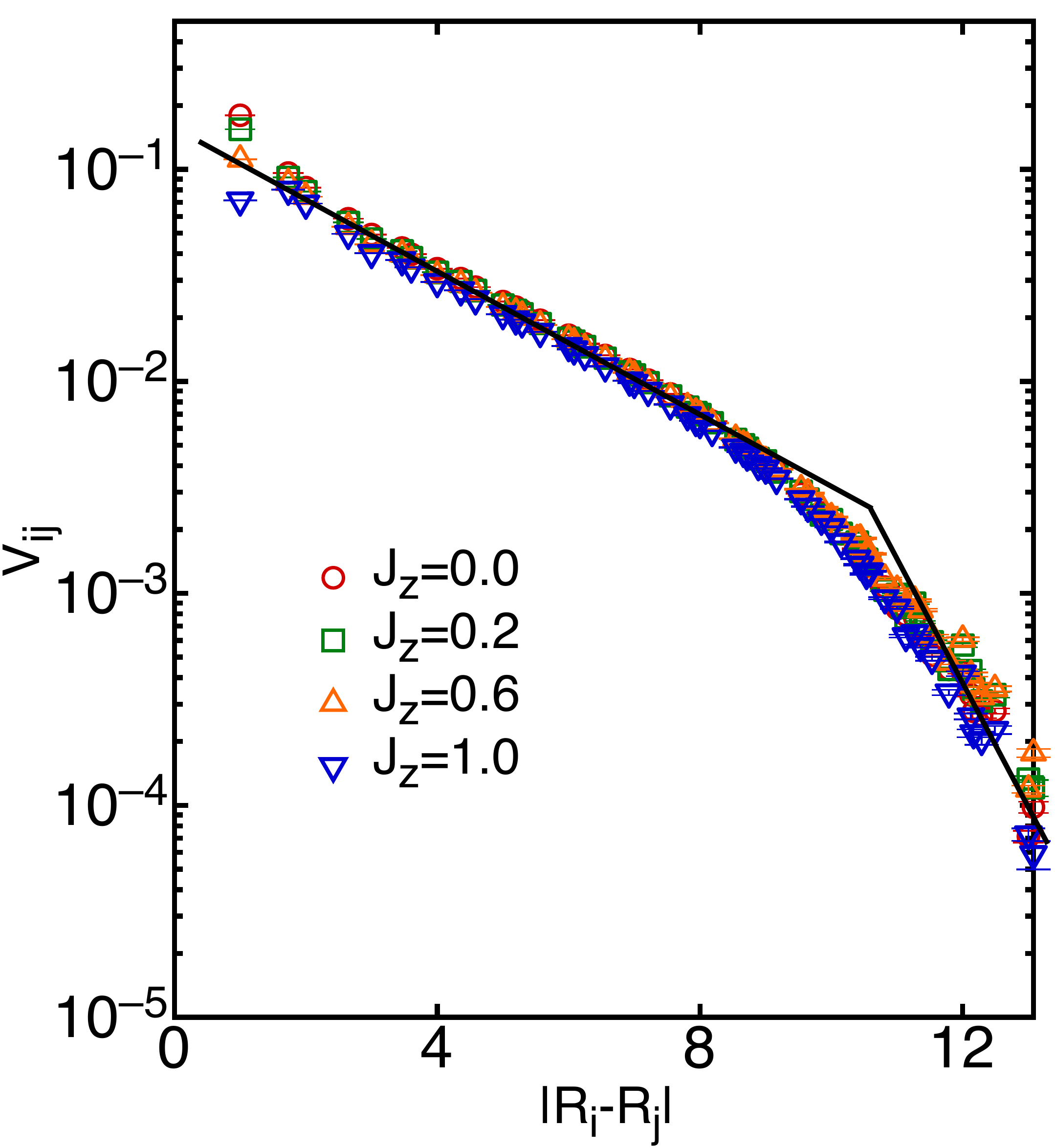}
\end{center}
\caption{\label{fig:jastrow}
(Color online) The parameters $v_{ij}$ as function of the distance $|i-j|$ for the $L=12$ lattice at 
$J_z=0$, $0.2$, $0.6$, and $1.0$. The black lines are guides for eye.}
\end{figure}
The inclusion of the spin Jastrow factor $\mathcal{J}_{s}={\rm exp}(1/2 \sum_{ij}v_{ij}S^{z}_{i}S^{z}_{j})$ 
gives rise to a considerable improvement of the $U(1)$ Dirac state in the whole region $0\le J_z\le 1$,
as we discussed in the main text. In Fig.~\ref{fig:jastrow}, we show the optimized parameters $v_{ij}$ as 
function of the distance $|R_i-R_j|$ between the sites $i$ and $j$. All parameters are positive and decay 
exponentially with distance, indicating that the spin Jastrow factor is short range. Moreover, the rate of 
decay seems to increase for $|R_i-R_j|>9$.

{\it The Lanczos steps--}
In the main part of the paper, we concentrate on variational wave functions as defined by
Eq.~(2); moreover, in few cases, we also improve them by applying a number $p$ of 
Lanczos steps:
\begin{equation}\label{eq:lanczos}
|\Psi_p\rangle = \left ( 1 + \sum_{m=1}^{p} \alpha_m H^m \right ) |\Psi_v\rangle,
\end{equation}
where $\alpha_m$ are $p$ additional variational parameters. Clearly, whenever $|\Psi_v\rangle$ 
is not orthogonal to the exact ground state, $|\Psi_p\rangle$ converges to it for large $p$. 
Unfortunately, on large sizes, only few steps can be efficiently afforded: here, we consider 
the case with $p=1$ and $p=2$ ($p=0$ corresponds to the original variational wave function).
Furthermore, an estimate of the exact energy may be obtained by the variance 
extrapolation. Indeed, for a systematically convergent sequence of states $|\Psi_p\rangle$ 
with energy $E_p$ and variance $\sigma_p^2$, it is easy to prove that 
$E_p \approx E_{\rm ex}+{\rm const} \times \sigma_p^2$, where 
$E_p=\langle \Psi_p|H|\Psi_p\rangle/N$ and
$\sigma_p^2=(\langle \Psi_p|H^2|\Psi_p\rangle-\langle \Psi_p|H|\Psi_p\rangle^2)/N$ are the 
energy and variance per site, respectively. Therefore, the exact energy $E_{\rm ex}$ may be 
extracted by fitting $E_p$ vs $\sigma_p^2$, for $p=0,1$, and $2$. 

\begin{figure*}
\begin{center}
\includegraphics[width=1.8\columnwidth]{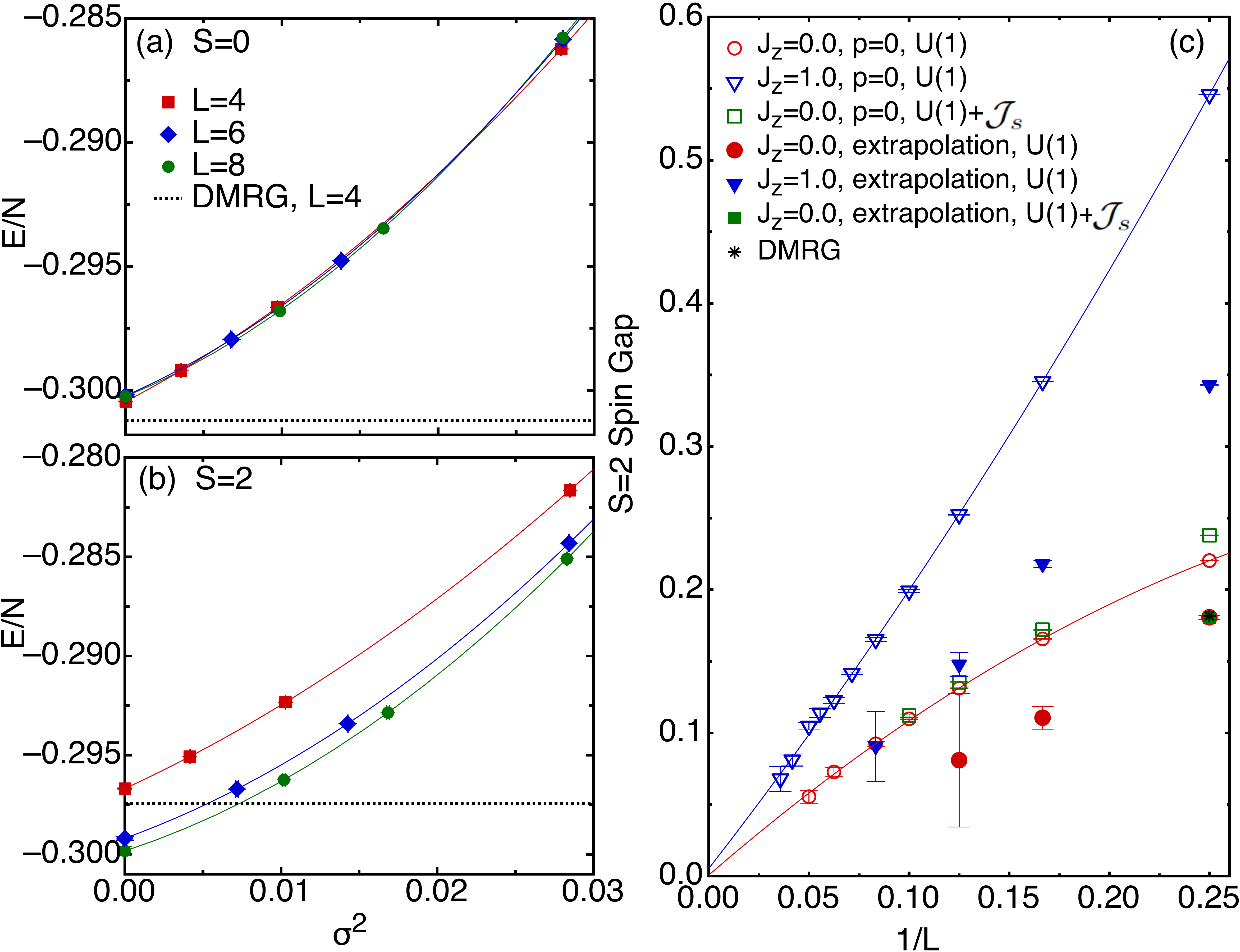}
\end{center}
\caption{\label{fig:extrapolation}
(Color online) Energies per site for the $S=0$ ground state (a) and $S=2$ excitation (b) 
versus the variance for $J_{z}=0$. The results with $p=0$, $1$, and $2$ are reported for 
$L=4$, $6$, and $8$. The variance extrapolated results are shown. The DMRG results 
on long cylinder with $L=4$ are also reported. 
(c) The $S=2$ spin gap with $p=0$ and extrapolation as a function 
of the inverse geometrical diameter ($1/L$) at $J_z=0$. The $U(1)$ Dirac state with and 
without Jastrow factor are both considered. The results for $J_{z}=1$ are from 
Ref.~\onlinecite{Iqbal2014}. On $L=4$ cluster, the results for the spin gap obtained by 
different wave functions are the same and equal to the one obtained by DMRG. For the $XY$ 
model, the upper bound of the $S=2$ spin gap is $0.05$ in the thermodynamic limit: this 
is entirely due to the large statistical error on $L=8$.}
\end{figure*}

Few Lanczos steps may be applied to the variational states with $S=0$ and $S=2$,
as described in Eq.~(\ref{eq:lanczos}), allowing a zero-variance extrapolation of the 
energies. Since this procedure is quite computational demanding, we only consider the $XY$
limit. Let us start by discussing the results on a small system with $L=4$.
For $S=0$, the $p=2$ state has $E=-0.299204(2)$, 
while, by performing the zero-variance extrapolation, the estimated 
energy is $E=-0.30045(1)$, which is quite close to the DMRG result $E=-0.301228$ 
on long cylinder (see Fig.~\ref{fig:extrapolation}a). For the $S=2$ excitation, the $p=2$ energy is 
$E=-0.295072(2)$ and the zero-variance extrapolation gives $E=-0.29668(3)$, again very
close to the DMRG result of $E=-0.29744$ (see Fig.~\ref{fig:extrapolation}b). Remarkably, 
the extrapolated gap that we obtain agrees with the DMRG one, indicating that there is 
an almost exact cancellation error between the $S=0$ and $S=2$ energies. 

All energies with $p=0$, $1$, and $2$ Lanczos steps and their zero-variance extrapolation 
on $L=4$, $6$, and $8$ clusters are reported in Table~\ref{lsdata} and 
Fig.~\ref{fig:extrapolation}(a and b). Compared to the Heisenberg model ($J_{z}=1$),~\cite{Iqbal2014}
the $S=2$ spin gaps on finite size clusters ($L=4$, $6$, and $8$) are smaller 
(Fig.~\ref{fig:extrapolation}(c)). This is consistent with the DMRG calculations.~\cite{Wei2014}
However, the variance in the $XY$ model is almost twice time larger than the one in the 
Heisenberg model for each Lanczos step. This fact indicates that the $U(1)$ Dirac state 
is less accurate to describe the ground state when decreasing $J_{z}$. Nevertheless, the 
thermodynamic extrapolation of the $S=2$ gap is still possible, see Fig.~\ref{fig:extrapolation}(c). 
Here, the large error bar of the $L=8$ cluster is entirely due to the large variances of
the $S=0$ and $S=2$ states, which makes a rather imprecise extrapolation of zero
variance. By performing a fit of the three sizes with $L=4$, $6$ and $8$, which takes 
into account their error bar, we obtain a vanishing $S=2$ spin gap in the thermodynamic 
limit. Taking into account all the statistical errors of the fitting procedure, the 
largest possible value for the thermodynamic gap is about $0.05$.

In the main text we have shown that, the spin Jastrow factor improves the ground-state energy 
of the pure $U(1)$ Dirac state. We have also performed the Lanczos steps on this wave function
for the $XY$ model, and obtained the smaller variance with $p=0$ and $1$. However, the $p=2$ 
calculation is unstable: with a small change in the $p=2$ Lanczos parameters, the variance may
have large variations, while the energy does not change much. This fact may indicate that
there are many low-lying states with competing energy for the $XY$ model. Nevertheless, on the $L=4$ 
cluster, by performing a linear extrapolation with $p=0$ and $1$ results, we get the DMRG 
energies within one error bar for both the ground state and $S=2$ excitation.

\begin{table*}
\centering
\caption{\label{lsdata}
Energies of the $U(1)$ Dirac spin liquid (columns $2$-$5$) and its $S=2$ excitation 
(columns $6$-$9$), with $p=0$, $1$, and $2$ Lanczos steps on different clusters 
for the spin-$1/2$ $XY$ model. The estimated energies of the $S=0$ and $S=2$ 
states by using the zero-variance extrapolation marked in bold.}
\begin{tabular}{c|cccc|cccc}
 \hline \hline
       \multicolumn{1}{c|}{}
    & \multicolumn{1}{c}{$p=0$}
    & \multicolumn{1}{c}{$p=1$} 
    & \multicolumn{1}{c}{$p=2$}
    & \multicolumn{1}{c|}{$S=0$}
    & \multicolumn{1}{c}{$p=0$}
    & \multicolumn{1}{c}{$p=1$}
    & \multicolumn{1}{c}{$p=2$}  
    & \multicolumn{1}{c}{$S=2$}  \\ \hline       
$L=4$ & $-0.2862336(7)$ & $-0.2966459(4)$ & $-0.299204(2)$ & $\bm{-0.30045(1)}$ & $-0.2816427(7)$ & $-0.2923336(5)$ & $-0.295072(2)$ & $\bm{-0.29668(3)}$   \\                                         
$L=6$ & $-0.2858440(5)$ & $-0.2947757(7)$ & $-0.297948(1)$ & $\bm{-0.30024(5)}$ & $-0.2843146(5)$ & $-0.2934122(6)$ & $-0.296711(2)$ & $\bm{-0.29921(5)}$   \\                                                         
$L=8$ & $-0.2857821(6)$ & $-0.2934697(6)$ & $-0.296803(3)$ & $\bm{-0.3002(1)}$ & $-0.2850986(5)$ & $-0.292854(1)$ & $-0.296242(3)$ & $\bm{-0.2998(2)}$   \\
\hline \hline
\end{tabular}
\end{table*}

\end{document}